\documentclass[aps,pra,twocolumn,notitlepage,superscriptaddress,showpacs,10pt,longbibliography,floatfix]{revtex4-2} 

\usepackage{amsmath}    
\usepackage{amsfonts}
\usepackage{bm}
\usepackage{amssymb}
\usepackage{graphicx}   
\usepackage[usenames,dvipsnames]{xcolor}      
\usepackage{comment}
\usepackage{siunitx}
\usepackage{enumitem}
\usepackage{upgreek}
\usepackage{color}
\usepackage{marvosym,amsthm}

\usepackage{soul}
\usepackage[normalem]{ulem}

\mathchardef\mhyphen="2D

\usepackage{xr}
\makeatletter

\newcommand*{\addFileDependency}[1]{
\typeout{(#1)}
\@addtofilelist{#1}
\IfFileExists{#1}{}{\typeout{No file #1.}}
}\makeatother

\newcommand*{\myexternaldocument}[1]{%
\externaldocument{#1}%
\addFileDependency{#1.tex}%
\addFileDependency{#1.aux}%
}

\myexternaldocument{supp}

\begin{document}

\title{Topological routing in Chern insulators}

\author{Mark J. Ablowitz}
\affiliation{Department of Applied Mathematics, University of Colorado, Boulder, Colorado 80309, USA}

\author{Justin T. Cole}
\email[]{jcole13@uccs.edu}
\affiliation{Department of Mathematics, University of Colorado, Colorado Springs, Colorado 80918, USA}

\author{Sean D. Nixon}
\affiliation{Department of Applied Mathematics, University of Colorado, Boulder, Colorado 80309, USA}

\date{\today}

\begin{abstract}

Chern insulator systems are realizable in numerous physical systems and can support robust non-reciprocal transmission of energy.
A routing functionality constructed from two counter-oriented Chern insulator regions, using coupled Haldane type systems is proposed.  By adjusting the strength of a magnetic field and the frequency of an antenna source, it possible to steer 
the flow of energy: completely to the left, completely to the right, or split. Alternatively, two sources can be used 
to direct the flow of energy. This formulation has the potential to serve as a robust and reconfigurable component in optical transmission.

\end{abstract}

\maketitle

\section{Introduction}
\label{sec_intro}

Over the last four decades topological insulators have been investigated in numerous physical systems  \cite{Ozawa2019,Hasan2010,Ablowitz2022,Wang2009,Poo2011}. From their quantum Hall effect origins, topological insulators have been characterized by  special topological invariants that distinguish topologically protected states \cite{Klitzing1980,Haldane1988,Thouless1982}. Associated with these nontrivial invariants are topologically protected modes characterized by exceptional robustness. In particular, Chern insulators possess chiral edge states that are effective conductors along the boundary of topological  media. Furthermore, these systems are nonreciprocal: energy propagates in a preferred chiral direction.

While several fundamental features of Chern insulators are well-known, how they could be utilized in  applications is an important research direction. The  non-reciprocal nature of these systems makes them ideal candidates for  isolators \cite{Zhou2017,ElGanainy2015}, circulators \cite{Zhang2021,Nagulu2022},  and current dividers \cite{Ovchinnikov2022,Ding2025}, among others. Furthermore, the topological protection of these systems makes them robust to fabrication defects.

The purpose of this work is to introduce a switching feature that is capable to routing energy in a nonlocal fashion. Antennas, located well away from the switching junction, are used to control and steer  the flow of energy. Aspects of this approach have 
 been discussed in the context of a Floquet insulator with helically rotating waveguides  \cite{Ablowitz2024}. Unlike that work, here we: (a) introduce the effect in a non-Floquet model and (b) control the switching via antenna sources rather than through  carefully constructed eigenmodes. 
 
 A switching  mechanism has  been experimentally observed  in a magneto-optical lattice \cite{Tang2024}. Another switching effect in a magneto-optical configuration  was explored in \cite{Skirlo2014} where a routing was controlled by effectively changing the length of the interface through a defect barrier. Dynamic interfaces have also been explored in polariton \cite{Wong2026} and micro-ring resonator \cite{Dai2024} constructions for the purpose of routing. The fixed nature of our interface boundary differs from those. The different physical basis of these works highlight the generic nature of this switching mechanism in Chern insulator systems.  As a result, in this work we consider a generic minimal model which is motivated by Chern insulators and exhibits similar routing properties.

Specifically, we construct two counter-oriented chiral regions which induce a topological interface between hexagonal lattices; see  Fig.~\ref{NNN_interactions}. 
The system we analyze is modeled by a set of Haldane equations  \cite{Haldane1988} that differ in the orientation of their local fluxes. Notably, this system supports two topological interface modes and two degrees of freedom.   A Floquet version of this formulation was considered in \cite{Ablowitz2024,Shi2021}. We note that a Haldane interface model separating a trivial and non-trivial regions was considered in \cite{Goldman2016}, which only supported one chiral state on each interface.

As a wave propagating along the interface  encounters the end junction, it must turn left or right; there is no backscatter. By judiciously tuning the field strength or source frequency, it is possible to dictate the flow of the light. Namely, it is possible to steer light completely left, completely right, or a predetermined splitting of energy. Another approach we investigate interferes two source antennas to control the energy flow. The  benefit of the latter approach is that, in general, it is possible to  steer the light without changing the physical properties of the system. These ideas allow the precise and robust routing of energy, e.g. electromagnetic waves.

 Lastly, we remark that the Haldane model is a universal Chern insulator model and has been derived in a variety of systems, such as: ultracold fermionic systems \cite{Jotzu2014}, electronic gases \cite{Lannebere2018}, Floquet photonic waveguides \cite{Ablowitz2023} and magneto-optical lattices \cite{Ablowitz2024a}. As a result, our results find application in a wide variety physical systems.

\section{Interface Tight-binding model}
\label{model_sec}

The Haldane model is a quintessential model for Chern insulator honeycomb systems. When time-reversal symmetry is broken 
the Haldane model supports topologically protected chiral states that  are robust to lattice defects. Depending on the orientation of the pseudo-magnetic field, two-dimensional chiral states may propagate either clockwise or counterclockwise along the domain boundaries. 
Here we examine 
a tight-binding model that couples two counter-oriented domains and induces an interface.

\begin{figure}
\centering
   \includegraphics[scale = 0.32]{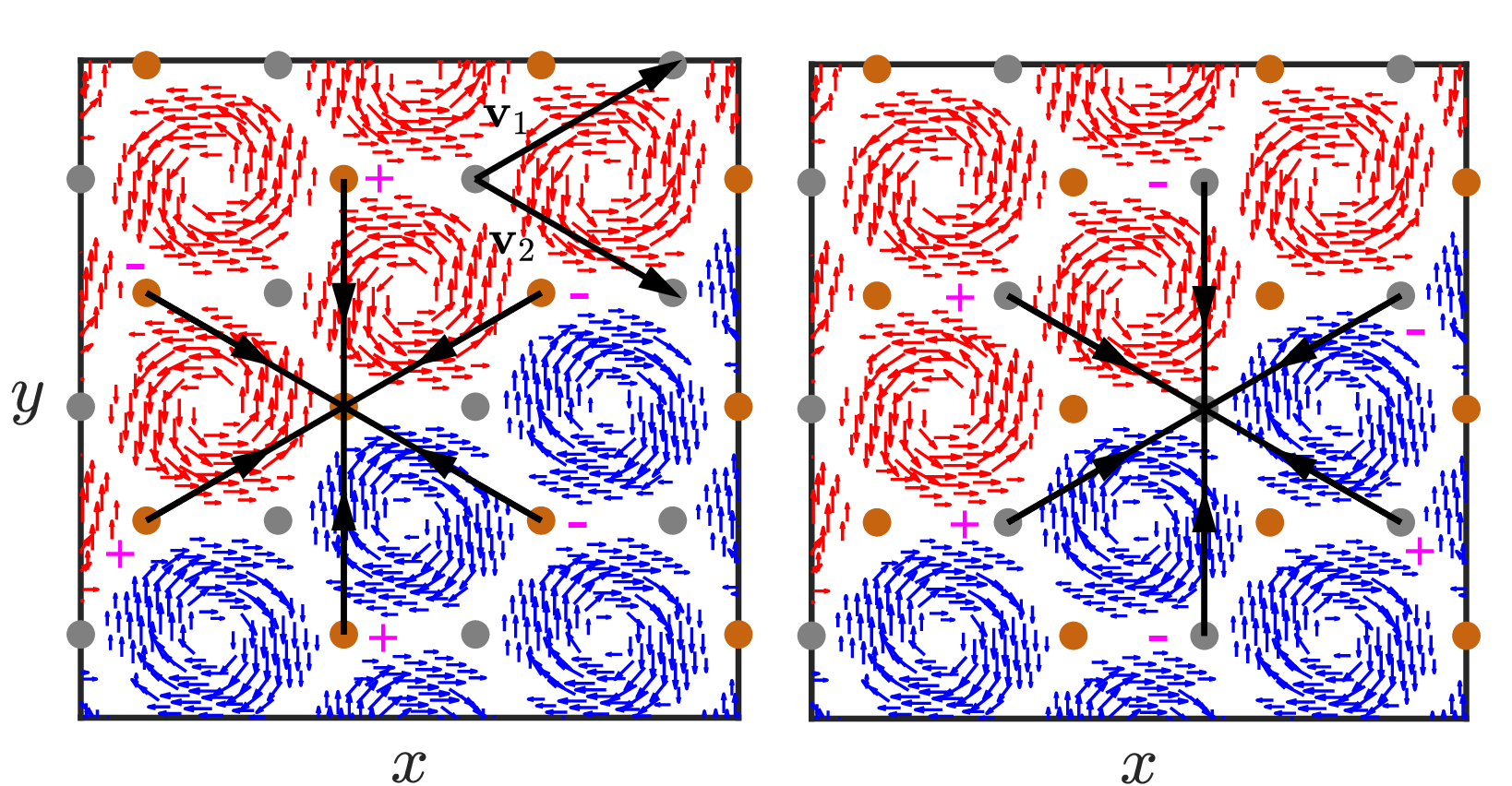}
    \caption{A zig-zag interface created by two oppositely oriented chiral regions. The periodic vector potential ${\bf A} ({\bf r})$ is locally counterclockwise (clockwise)   for $n <0~(n> 0)$, denoted by the red (blue) vector field. Black arrows indicate magnetic flux paths  between next-nearest neighbors for (left) a-sites and (right) b-sites. Magenta $\pm$ values indicate the sign of the phase for this interaction.
 \label{NNN_interactions}}
\end{figure}

Consider a lattice array consisting of orbital centers (lattice sites) arranged in a honeycomb configuration  (see Fig.~\ref{NNN_interactions}) with lattice vectors
$ {\bf v}_1 = ( 3/2 , \sqrt{3}/2 ) $ and ${\bf v}_2 = (3/2 , - \sqrt{3}/2), $
having nondimensionalized length $|| {\bf v}_{1,2}  || = \sqrt{3}$. 
The lattice cell $(m,n)$ is positioned at ${\bf v} = m {\bf v}_1 + n {\bf v}_2$  where $m,n \in \mathbb{Z}$. 
The corresponding reciprocal lattice vectors, 
$ {\bf k}_1 = 2 \pi  ( 1/3 ,  1/ \sqrt{3} ) $ and $ {\bf k}_2 =  2 \pi (1/3 , - 1/\sqrt{3} ), $
are chosen such that ${\bf k}_i \cdot {\bf v}_j = \delta_{ij}$.

Next, as in  \cite{Haldane1988}, consider a planar vector potential ${\bf A}(x,y)$  that  shares the periodicity of the  lattice. Furthermore, suppose the corresponding flux density ${\bf B} = \nabla \times {\bf A}$ yields a net zero flux over each hexagonal cell; that is, $\iint_S {\bf B} \cdot d{\bf S} = 0$. The  next-nearest neighbor coefficients  are complex-valued and have  the polar form $t_2 e^{i \phi}, t_2 \ge 0$, where the phase $\phi$ is proportional to the local flux $\int {\bf A} \cdot d{\bf r}$ along the path between next-nearest neighbors.

A model for a vector potential that induces such an interface is given by \cite{Lannebere2018}
\begin{align}
\nonumber
{\bf A}({\bf r}) = & \alpha ~{\rm sgn}(n)   \big[  {\bf k}_1  \sin \left( {\bf k}_1 \cdot {\bf R} \right) +  {\bf k}_2  \sin \left( {\bf k}_2 \cdot {\bf R} \right)  \\ \label{vec_pot}
&+  ({\bf k}_1 + {\bf k}_2)  \sin \left(( {\bf k}_1 + {\bf k}_2)  \cdot {\bf R} \right) \big] \times \hat{\bf z} ,
\end{align}
where ${\bf R} = {\bf r} - {\bf r}_c$ is a shifted coordinate by ${\bf r}_c$ and $\alpha > 0$ is the field strength. By adjusting the direction of the field (through the sgn term), it is possible to engineer an interface that supports localized states. 
In particular, the interface induced by (\ref{vec_pot})  is located at $
n =0$ and shown in Fig.~\ref{NNN_interactions}.

 A nontrivial flux is acquired when the path between two lattices sites intercepts the vector potential. That is, there is  a positive (negative) flux effect when it moves with (against)  the local flux; see magenta labels in Fig.~\ref{NNN_interactions}. 
  As the field is constructed, there is no significant flux accumulation among nearest neighbor interactions implying completely real coefficients.

Now we describe our Haldane interface model. Consider two  oppositely oriented Chern insulator slabs placed side-by-side  along a zig-zag interface (see Fig.~\ref{NNN_interactions}), parallel to the ${\bf v}_1$ lattice vector. The parameters are chosen so that the top-left (bottom-right) region supports  counterclockwise (clockwise) chiral motion.  Away from the interface,  the systems behave  like standard Haldane models with their respective fluxes. On the other hand, at the interface next-nearest terms must cross the boundary and interact with  lattice sites on the other side. As a result of this configuration, there is a net flow in the $+ {\bf v}_1$ direction. Moreover, the system is Hermitian.
 Lastly,  the a-sites and b-sites experience self-interaction detuning given by  $+\mathcal{M}$ and $- \mathcal{M}$, respectively, corresponding to inversion symmetry breaking when $ \mathcal{M}\not=0$. 

The governing system of equations describing time-harmonic solutions of this system can be compactly expressed as the following eigenvalue problem in terms of the eigenvalue problem
\begin{equation}
\label{compact_time_ind}
\lambda {\bf v} = H {\bf v} ,
\end{equation}
 where ${\bf v} = [{\bf a} | {\bf b} ]^T$ is the eigenvector with a-sites and b-sites and  $H$ is the Hermian matrix describing the interface system in Eqs.~(\ref{a_eqn})-(\ref{b_int_eqn}). The explicit representation of this matrix is given in  Appendix~\ref{Hamil_define_sec}.

Away from the interface, the tight-binding model 
 is described by the system of equations \\
 ${\bf n\not=0:}$
\begin{align}
 \label{a_eqn}
\lambda a_{mn}  & =  \mathcal{M} a_{mn} + t_1 \left( b_{mn} + b_{m-1,n} + b_{m,n-1} \right) \\ \nonumber
& + t_2 e^{- i   {\rm sgn}(n) \phi } \left( a_{m,n+1} + a_{m-1,n} + a_{m+1,n-1} \right) \\ \nonumber
& + t_2 e^{ i   {\rm sgn}(n) \phi } \left( a_{m+1,n} + a_{m,n-1} + a_{m-1,n+1} \right) , \\ \nonumber \\  \label{b_eqn}
\lambda b_{mn}  & =  -\mathcal{M} b_{mn} + t_1 \left( a_{mn} + a_{m+1,n} + a_{m,n+1} \right) \\ \nonumber
& + t_2 e^{ i  {\rm sgn}(n)  \phi } \left( b_{m,n+1} + b_{m-1,n} + b_{m+1,n-1} \right) \\ \nonumber
& + t_2 e^{- i  {\rm sgn}(n) \phi } \left( b_{m+1,n} + b_{m,n-1} + b_{m-1,n+1} \right) ,
\end{align}
where $\lambda,\mathcal{M},t_1,t_2 ,\phi$ are constant and the signs
correspond to  the up-and-left ($n < 0$) and down-and-right ($n>0$) regions, respectively. Notice that this is simply two copies of the Haldane model with oppositely oriented flux.
The governing equations along the interface are given by \\
 $\bf{n=0:}$
\begin{align}
 \label{a_int_eqn}
\lambda a_{m,0}  & =  \mathcal{M} a_{m,0} + t_1 \left( b_{m,0} + b_{m-1,0} + b_{m,-1} \right) \\ \nonumber
& + t_2 e^{ i  \phi } \left(   a_{m-1,1} + a_{m-1,0}  +  a_{m+1,-1}  \right) \\ \nonumber
& + t_2 e^{- i \phi } \left( a_{m+1,0} + a_{m,-1} + a_{m,1}  \right) , \\ \nonumber \\ \label{b_int_eqn}
\lambda b_{m,0}  & =  -\mathcal{M} b_{m,0} + t_1 \left( a_{m,0} + a_{m+1,0} + a_{m,1} \right) \\  \nonumber
& + t_2 e^{- i   \phi } \left( b_{m+1,0} + b_{m-1,1}  + b_{m+1,-1}  \right) \\ \nonumber
& + t_2 e^{ i   \phi } \left(  b_{m,1}  + b_{m,-1}  + b_{m-1,0} \right) .
\end{align}
At first glance, Eqs.~(\ref{a_int_eqn})-(\ref{b_int_eqn}) look similar to (\ref{a_eqn})-(\ref{b_eqn}); there is no change in the self and nearest neighbor  couplings. The only difference is the next-nearest neighbor interactions that cross the interface; their flux does not completely match those in the bulk regions. The signs of the phases can be inferred from the diagram in  Fig.~\ref{NNN_interactions}.

\section{Semi-infinite  Strip}
\label{interface_mode_sec}

\begin{figure}
\centering
   \includegraphics[scale = 0.28]{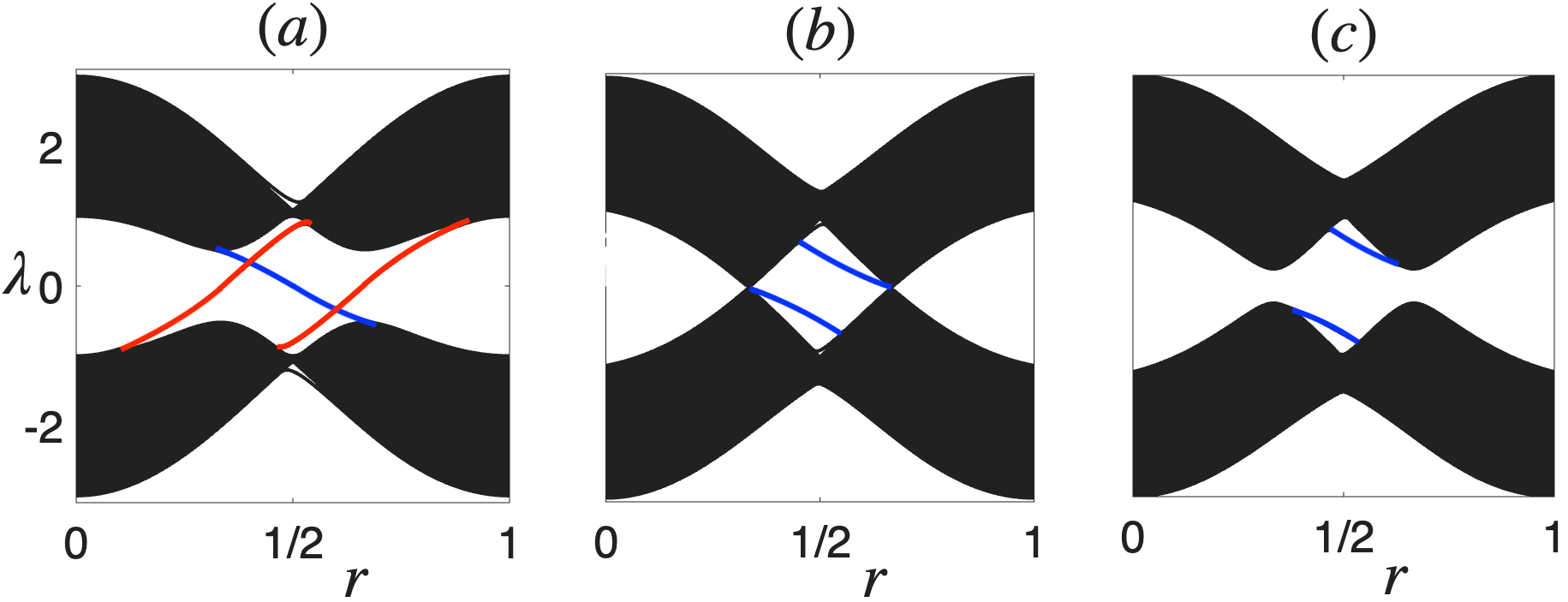}
    \caption{Spectral edge bands
     for the interface-strip problem using parameters $t_1 = 1, t_2 = 0.1, \phi = \pi/2, N = 128$ and (a) $ \mathcal{M} = 0$, (b) $ \mathcal{M} = - 3 \sqrt{3} t_2 \sin \phi$, and (c) $  \mathcal{M} = - 3 \sqrt{3} t_2 \sin \phi - 1/4$. A topological transition takes place between $(a)$ and $(c) $. Red (blue) bands denote interface (edge) modes. 
 \label{band_diag_compare}}
\end{figure}

In this section localized interface modes and their spectral bands are examined. 
To begin, consider a semi-infinite lattice strip: infinite in the ${\bf v}_1$-direction (parallel to the interface) and finite with open (zero) boundary conditions in the ${\bf v}_2$-direction 
$$a_{m,\pm (N+1)} = 0, ~~~ b_{m,\pm (N+1)} = 0 , $$
for a  positive integer $N$. Lattice sites in the region  $-N \le n < 0$  and $0 < n \le N$ correspond to the left and right domains, respectively.
 Look for Fourier solutions of the form 
$$ \begin{pmatrix} a_{mn} \\ b_{mn} \end{pmatrix} =  \int e^{i  {\bf k} \cdot m {\bf v}_1 } \begin{pmatrix} a_{n}({\bf k})  \\ b_{n}({\bf k})  \end{pmatrix} d{\bf k} ,$$
such that ${\bf k} \cdot {\bf v}_1 = (r {\bf k}_1 + s {\bf k}_2) \cdot {\bf v}_1 = 2 \pi r$ for $r \in [0,1]$ due to 
${\bf k}_i \cdot {\bf v}_j = 2\pi \delta_{ij}.$
This yields the one-dimensional reduction of system (\ref{a_eqn})-(\ref{b_eqn}), \\
 ${\bf n\not=0:}$
\begin{align}
 \label{a_eqn_1d}
\lambda(r) a_{n}  & =  \mathcal{M} a_{n} + t_1 \left( b_{n} + e^{-i 2 \pi  r} b_{n} + b_{n-1} \right) \\ \nonumber
& + t_2 e^{- i  {\rm sgn}(n) \phi } \left( a_{n+1} + e^{-i 2 \pi  r} a_{n} + e^{i 2 \pi  r} a_{n-1} \right) \\ \nonumber
& + t_2 e^{ i  {\rm sgn}(n)\phi } \left( e^{i 2 \pi  r} a_{n} + a_{n-1} + e^{-i 2 \pi  r} a_{n+1} \right) , \\ \nonumber \\  \label{b_eqn_1d}
\lambda(r) b_{n}  & =  - \mathcal{M} b_{n} + t_1 \left( a_{n} + e^{i 2 \pi  r} a_{n} + a_{n+1} \right) \\ \nonumber
& + t_2 e^{ i {\rm sgn}(n)  \phi } \left( b_{n+1} +e^{-i 2 \pi  r} b_{n} + e^{i 2 \pi  r} b_{n-1} \right) \\ \nonumber
& + t_2 e^{- i  {\rm sgn}(n) \phi } \left( e^{i 2 \pi r} b_{n} + b_{n-1} +e^{-i 2 \pi  r}  b_{n+1} \right) ,
\end{align}
and along the interface, the governing equations (\ref{a_int_eqn})-(\ref{b_int_eqn}) reduce to \\
 ${\bf n=0:}$
\begin{align}
 \label{a_int_eqn_1d}
\lambda(r) a_{0}  & =  \mathcal{M} a_{0} + t_1 \left( b_{0} +e^{-i 2 \pi  r}   b_{0} + b_{-1} \right) \\ \nonumber
& + t_2 e^{ i  \phi } \left(   e^{-i 2 \pi  r}  a_{1} + e^{-i 2 \pi  r}  a_{0}  +  e^{i 2 \pi  r}  a_{-1}  \right) \\ \nonumber
& + t_2 e^{- i \phi } \left( e^{i 2 \pi  r}  a_{0} + a_{-1} + a_{1}  \right) , \\ \nonumber \\ \label{b_int_eqn_1d}
\lambda(r) b_{0}  & =  - \mathcal{M} b_{0} + t_1 \left( a_{0} + e^{i 2 \pi  r}  a_{0} + a_{1} \right) \\  \nonumber
& + t_2 e^{- i   \phi } \left( e^{i 2 \pi  r}  b_{0} + e^{-i 2 \pi  r}  b_{1}  +e^{i 2 \pi r}   b_{-1}  \right) \\ \nonumber
& + t_2 e^{ i   \phi } \left(  b_{1}  + b_{-1}  +e^{-i 2 \pi  r}   b_{0} \right) .
\end{align}

\begin{figure}
\centering
   \includegraphics[scale = 0.35]{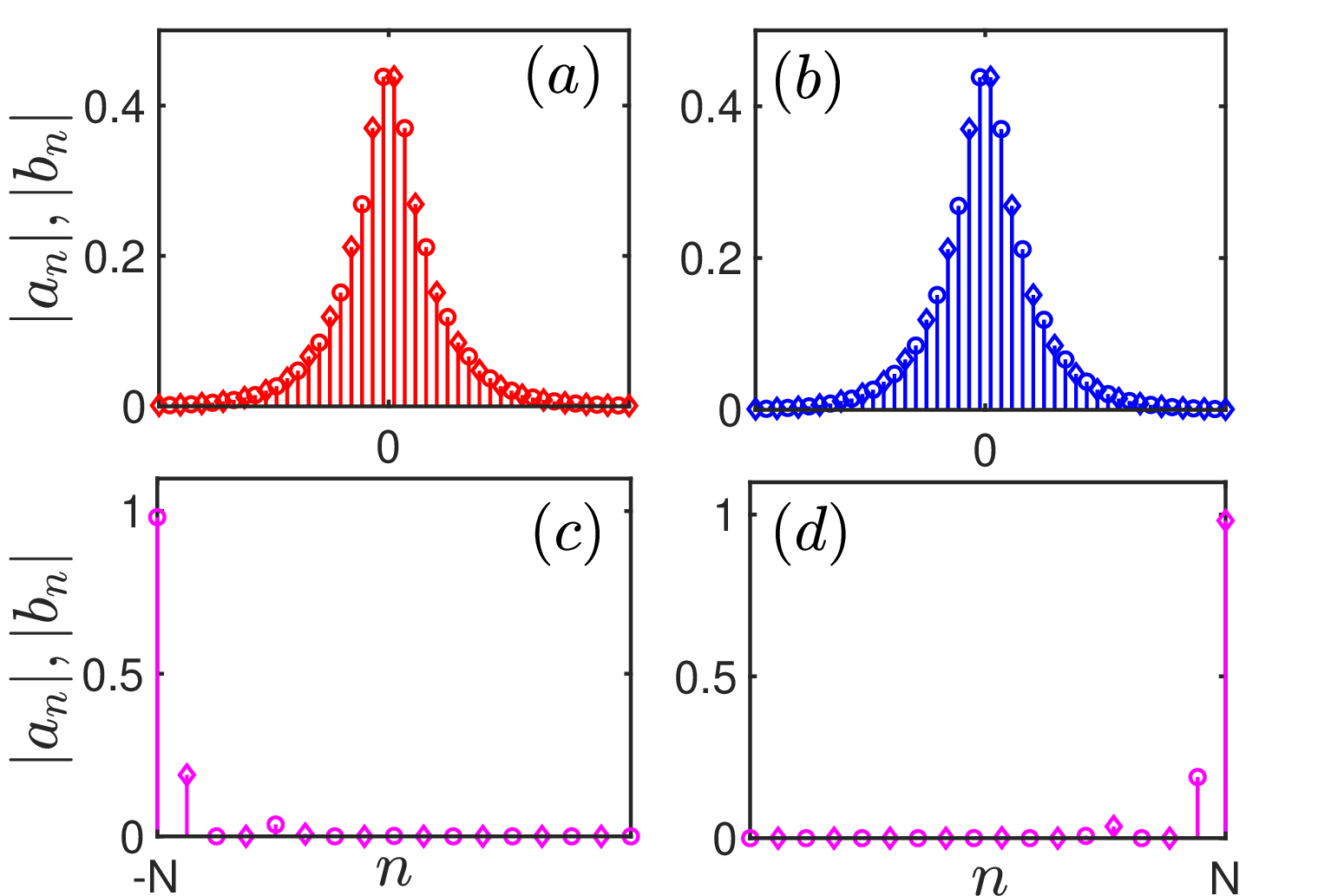}
    \caption{Eigenmodes
   corresponding to the edge 
    band diagram in Fig.~\ref{band_diag_compare}(a) at $\lambda = 0$. Interface modes correspond to (a) $r \approx 0.3381$, (b) $r \approx 0.6619$, and the edges modes (c),(d) are at $r = 0.5$. Circles (diamonds) denote $a$-sites ($b$-sites).
 \label{eigenmode_plot}}
\end{figure}

Edge band 
diagrams for a wide strip are shown in Fig.~\ref{band_diag_compare}. 
Topologically nontrivial states are present when time-reversal symmetry is broken ($t_2 >0 , \phi \not= \pi \mathbb{Z}$) and there is or nearly is inversion symmetry ($\mathcal{M} \approx 0$). At the interface, there are two distinct families of localized modes, indicated by the red-colored curves. 
Additionally, there is a degeneracy of two blue curves corresponding modes 
localized along the outer edges (where the open boundary conditions are applied). A critical inversion value of  $\mathcal{M}_c \approx \pm 3 \sqrt{3} t_2 \sin \phi $ is highlighted in Fig.~\ref{band_diag_compare}(b) where the band gap closes; the same critical point occurs in the single domain Haldane problem. Beyond this value ($|\mathcal{M}| > |\mathcal{M}_c|$), no gapless localized
 states are observed. 

The zero frequency ($\lambda = 0$) interface eigenmodes are highlighted in the top row of Fig.~\ref{eigenmode_plot}. For these parameters, both interface states are observed to be even symmetric in their magnitude about the midpoint. However, this symmetry appears to be somewhat special. That is, we do not see this parity symmetry for $\lambda \not= 0$ interface modes.
In addition, there are two edge modes at $r = 1/2$ shown in the bottom row of Fig.~\ref{eigenmode_plot};
localized on the left and right outer edges of the system. 
All eigenmodes shown  
are observed to decay exponentially fast away from their respective interfaces.

Scrutinizing the interface modes in Figs.~\ref{eigenmode_plot}(a)-(b),  each solution exhibits the  asymptotic behavior 
\begin{equation}
a_n(s) , b_n(s)  \sim 
\begin{cases} s^n,  & n \rightarrow \infty  \\ 
(s^*)^{-n},  & n \rightarrow - \infty 
\end{cases} ,
\end{equation}
where  $s \in \mathbb{C}$ and  $|s| < 1$. This says that the interface state approximately consists  of two exponentially decaying tails. At $\lambda =0$, the tails  decay in magnitude at the same rate; the only difference is the conjugation of the exponential base. Comparing the two interface states in Fig.~\ref{eigenmode_plot}(a)-(b), the main difference (other than an arbitrary constants) is $s \rightarrow s^*$, opposite conjugation; they decay in magnitude at the same rate.

\section{2D Boundary value problem}
\label{BVP_sec}

In this section a fully two-dimensional interface problem is studied by solving the system of equations (\ref{a_eqn})-(\ref{b_int_eqn}). The open   boundary conditions imposed are 
\begin{equation}
\label{zero_BCs}
\begin{split}
a_{m,\pm(N+1)} = & ~0 = b_{m,\pm(N+1)} , \\
a_{\pm(M+1),n} = & ~0 = b_{\pm(M+1),n} .
\end{split}
\end{equation}
This is a boundary value problem with the interface located at $n = 0$. 
The resulting system consists of $ 2 \times (2M +1) \times (2N +1)$ equations: $(2M+1)N$ sites in both 
subregions plus an interface layer, and an extra factor of 2 due to the two sublattices. These boundary conditions result in all edges and interfaces being zig-zag.

\begin{figure}
\centering
   \includegraphics[scale = 0.35]{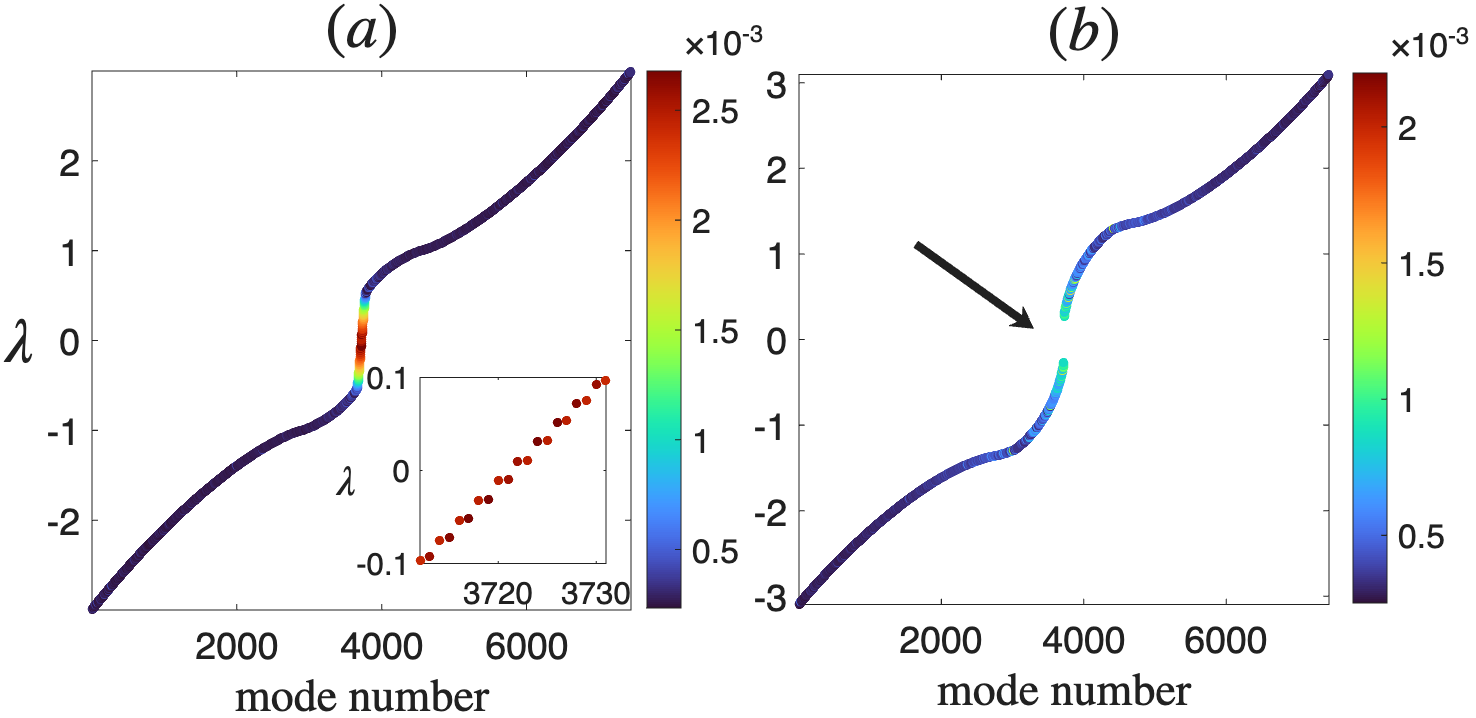}
    \caption{Eigenenergies for the 2D  interface problem (\ref{a_eqn})-(\ref{b_int_eqn}) with open boundary conditions and parameters: $t_1 = 1, t_2 = 0.1, \phi = \pi/2 $ with (a) $\mathcal{M} = 0$ and (b) $\mathcal{M} = -3\sqrt{3} t_2 \sin \phi - 1/4$. The color indicates the degree of localization as measured by the IPR (\ref{IPR}).
 \label{eigenenergy_2d_compare}}
\end{figure}

The full spectrum for a set of typical parameters is shown in Fig.~\ref{eigenenergy_2d_compare} using $M = N = 30$ sites in each direction.  
Two band configurations are shown: topological Fig.~\ref{eigenenergy_2d_compare}(a) and nontopological Fig.~\ref{eigenenergy_2d_compare}(b) (these topologies are established in the next section). Notably, the nontopological case has an energy gap near $\lambda = 0$ in which no localized modes are observed.

To quantify the degree of localization, we compute the inverse participation ratio (IPR)
\begin{equation}
\label{IPR}
{\rm IPR} = \sum_{m,n} | a_{mn} |^4 + | b_{mn} |^4 ,
\end{equation}
(the $l_2$-norm/power is fixed to 1). A  large (small) value of IPR indicates a relatively localized (delocalized) eigenmode. In Fig.~\ref{eigenenergy_2d_compare} we see that the most localized modes are near the zero energy. Indeed, the mid-gap eigenmodes are localized along the domain boundary (see next).

\begin{figure}
\centering
   \includegraphics[scale = 0.35]{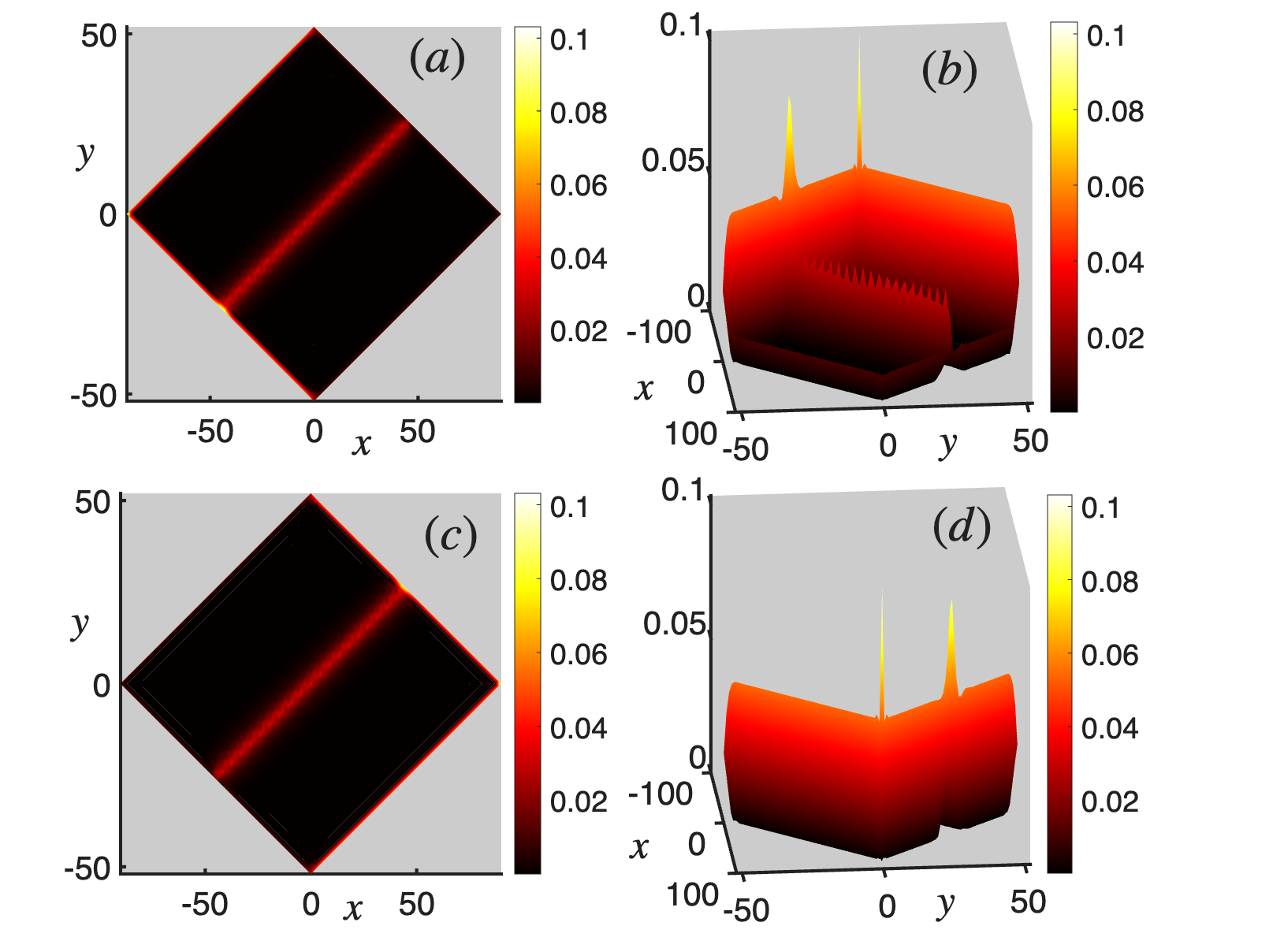}
    \caption{Interface eigenmodes corresponding to the parameters in Fig.~\ref{eigenenergy_2d_compare}(a) with  midgap eigenenergies $\lambda \approx \pm 0.009897$. Shown is a top view (left column) and  side view (right column) of (a),(b) $|a_{mn}|$  and (c),(d) $|b_{mn}|$. The most significant intensities occurs at the outermost sites.
 \label{eigenmodes_2d_plot}}
\end{figure}

A set of topological interface eigenmodes  corresponding to nearly
zero eigenfrequencies is shown in Fig.~\ref{eigenmodes_2d_plot}. Most of the energy is concentrated in the outermost row of lattice sites. For example,  if either $n = - N$ or $m = - M$ the a-sites are  outermost and the mode forms a ``wall" along the boundary of the domain; the process is similar for b-sites at $n = N$ or $m = M$. There is also a small strand of energy along the interface (see Figs.~\ref{eigenmodes_2d_plot}(a) and (c)). 

These states are important because they are   conductors of the chiral states in the next section. As such, it is essential that energy be able to conduct in, along, and out of  the interface. In particular, there are two ``exits'' from the interface: a ``left'' and a ``right'' set of paths. We will exploit this functionality to control the flow of energy and create an effective switching mechanism.

\subsection{Topology}

To establish the topology of this system, the spectral localizer is applied \cite{Cerjan2022,Cerjan2024,Wong2025}. Something like the localizer is essential since the multi-region configuration 
prevents a two-dimensional Fourier transform and subsequent analysis of bulk states and their topology. That is, establishing topology via direct application of the bulk-edge correspondence in a standard fashion is not possible. 

The spectral localizer is defined by  
\begin{align}
\label{spec_localizer}
&L(x,y,E) = \\ \nonumber
& \begin{pmatrix} H - E I & \kappa (X  - x I) - i \kappa (Y - y I) \\ 
 \kappa (X  - x I) + i \kappa (Y - y I)  & - (H - E I)  \end{pmatrix} ,
\end{align}
where $X,Y$ are the diagonal position  matrices consisting of the lattice site locations 
and $H$ is the full 2D matrix Hamiltonian operator of the interface problem given in Eq.~(\ref{compact_time_ind}). The inputs are the spatial coordinates $(x,y)$ and frequency $E$. The hyperparameter $\kappa$ is used to balance the spatial and frequency scales \cite{Cerjan2024a}. 
For the results shown here, we take the value $\kappa = 0 .1$; a typical value.

\begin{figure}
\centering
   \includegraphics[scale = 0.33]{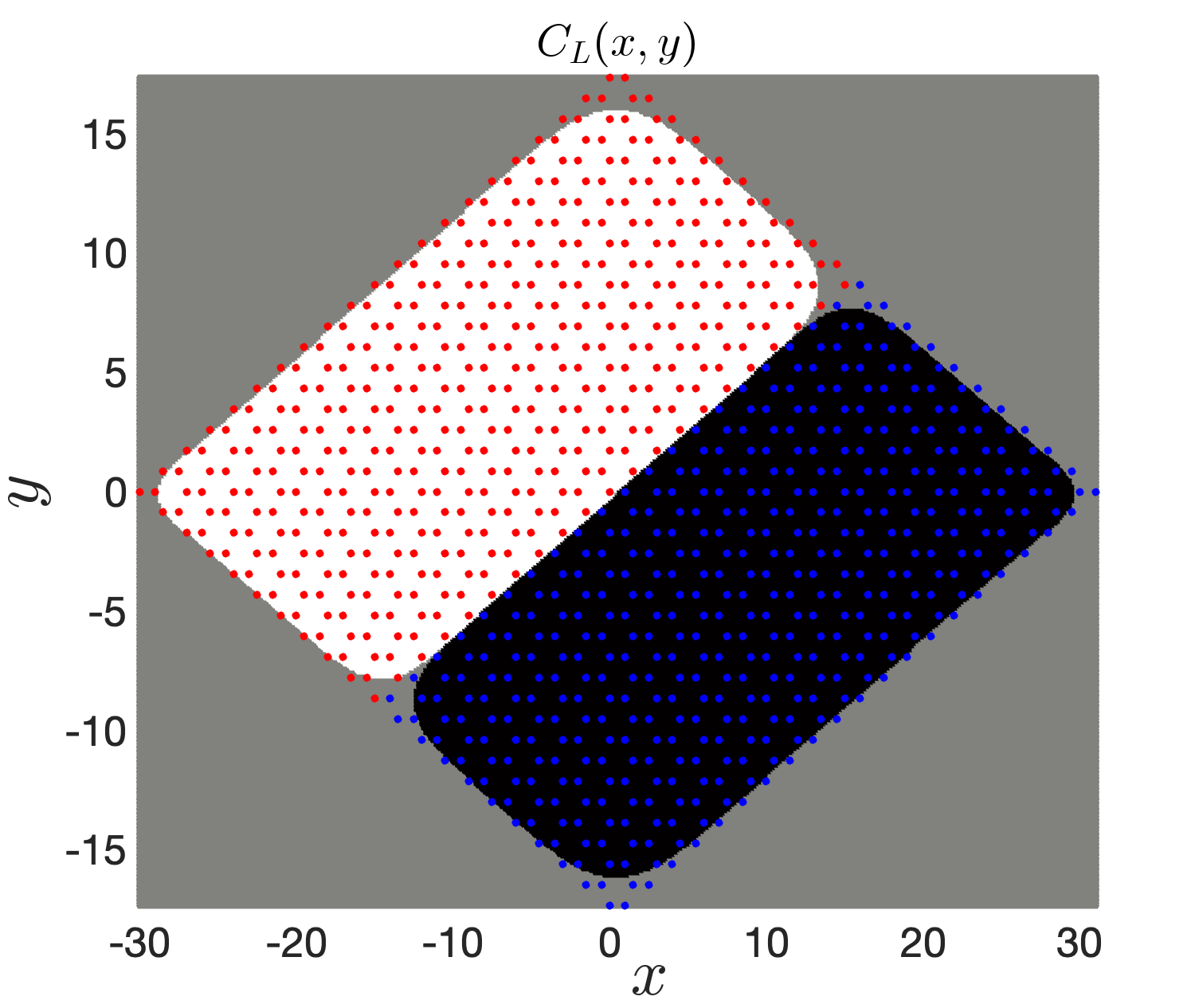}
    \caption{Local Chern number (\ref{local_chern_define}) values for the two regions at zero frequency. The  values are $C_L = +1$ (white region), $C_L = -1$ (black region), and $C_L = 0$ (grey region). Each region consists of a $10 \times 10$ honeycomb lattice using the same parameters as Fig.~\ref{eigenenergy_2d_compare}(a).  Lattice sites are denoted by red ($n < 0$) and blue  ($n > 0$) dots. For the trivial case in Fig.~\ref{eigenenergy_2d_compare}(b), the local Chern number is $C_L = 0$ everywhere. 
     \label{interface_topology}}
\end{figure}

To determine the topology at a given space-frequency location $(x,y,E)$, we compute the (real-space) local topology. Here we only study the zero energy modes, i.e. $E = 0$. The local Chern number at zero energy is defined by
\begin{equation}
\label{local_chern_define}
C_L(x,y) = \frac{1}{2} {\rm sig} \left[ L(x,y,0)  \right] , 
\end{equation}
where ${\rm sig}$ denotes the matrix signature: the number of positive eigenvalues minus the number of negative eigenvalues.

 A wide range of spatial values are scanned and local Chern number computed; the results are shown in Fig.~\ref{interface_topology}. The regions have Chern numbers of $\pm 1 $, respectively. As a result, the interface represents a change in Chern jump of $\Delta C = 2 $ and the presence of two chiral interface states. This agrees with the 1D bands shown in Fig.~\ref{band_diag_compare}(a). Lastly, for the parameters used in Fig.~\ref{band_diag_compare}(c) and Fig.~\ref{eigenenergy_2d_compare}(b), the local Chern number is zero everywhere; we do not show it here.

\begin{figure}
\centering
   \includegraphics[scale = 0.45]{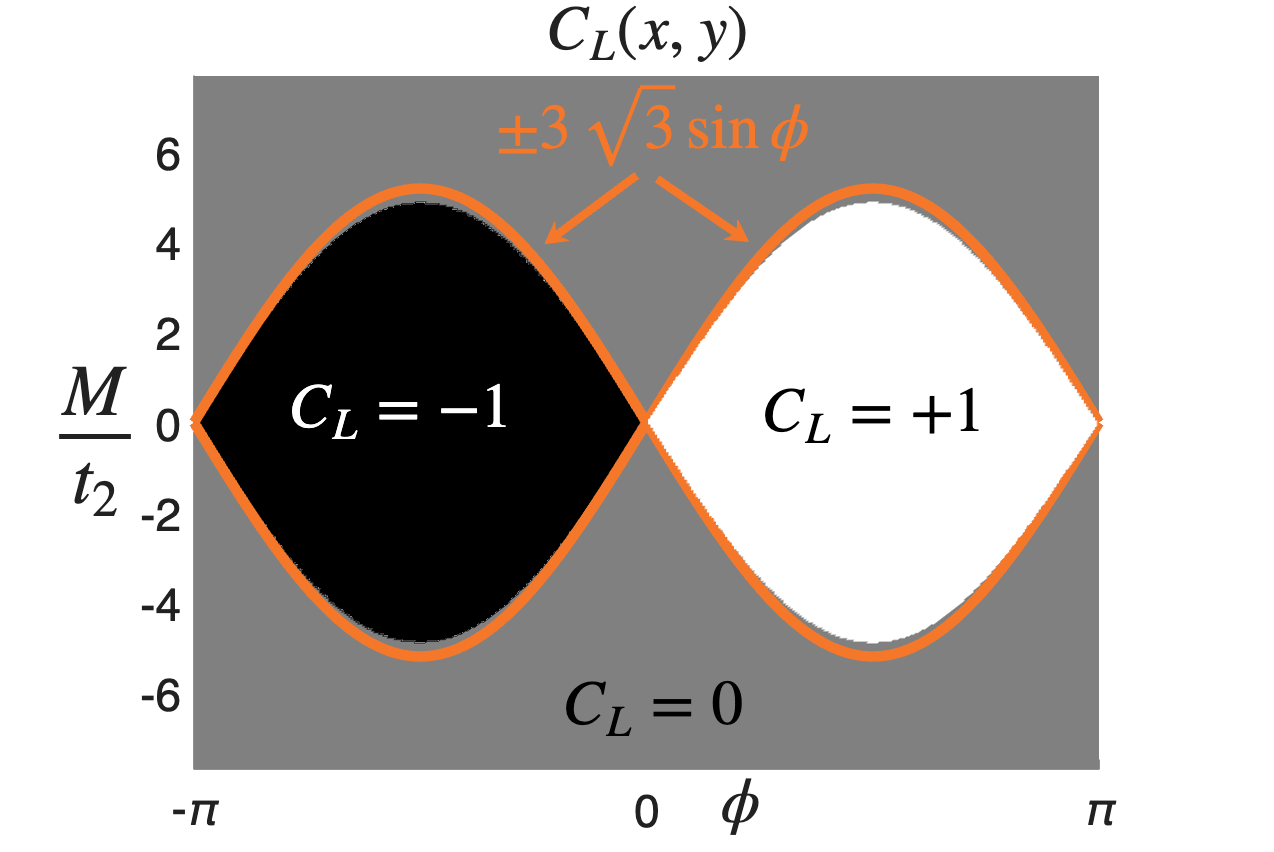}
    \caption{Local Chern number (\ref{local_chern_define}) values for an interior spatial point $(x_*,y_*) = (-6.5,4)$ at zero frequency $(E = 0)$ for the   $10 \times 10$ honeycomb lattice shown in Fig.~\ref{interface_topology}. For comparison, the bulk Chern topology transition curves are included.
     \label{top_phasediagram}}
\end{figure}

The topological dependence on the model parameters is highlighted in Fig.~\ref{top_phasediagram}. Here the local topology at a single spatial point $(x_*,y_*)$ located in the middle of the upper region is probed; the results are opposite in the lower region. The overall picture of topology is found to resemble that of the classic Haldane model; cf.  \cite{Haldane1988}.  In particular, sufficiently strong inversion symmetry breaking is found to lead to a topological transition, similar to what was observed in Fig.~\ref{band_diag_compare}, and changing the direction of the local flux results in oppositely-oriented chiral modes. 
For comparison, the Haldane topological transition (orange curve) is included in Fig.~\ref{top_phasediagram}. The small discrepancy between it and ours can, in part, be attributed to the finite lattice we are using. The orange curves are obtained from the bulk problem which assumes an effectively infinite region which does not feel boundary effects.

\section{Evolution along interface and routing}
\label{switch_sec}

 We now discuss the main result of this work: a robust and  topological switching effect. 
 To begin, we replace $\lambda \Leftrightarrow  i \partial_t $ in Eqs.~(\ref{a_eqn})-(\ref{b_int_eqn}) and take the inverse Fourier transform of the interface system.   The resulting system is integrated in time using a  fourth-order Runge-Kutta scheme.

The input is envisioned as a set of antennas that emit at a prescribed frequency. This construction  allows flexibility in adjusting the position and frequency of the field. To model this,  we consider the forced time-dependent system
\begin{equation}
\label{source_eqn}
 i \frac{d {\bf c}}{dt} =  H {\bf c}  + {\bf f}(t) ,
\end{equation}
where $H$ is the governing matrix in Eq.~(\ref{compact_time_ind})
that represents the solution of Eqs.~(\ref{a_eqn})-(\ref{b_int_eqn}) and ${\bf f}(t)$ models the antennas.
For all cases considered below, the initial state is zero, i.e. ${\bf c}(0) = {\bf 0}$. 

The antenna is constructed to excite a single lattice site at a prescribed
frequency. As such, it will take the form $[{\bf f}]_{mn} = \delta_{pm} \delta _{qn} f_{\lambda}(t)$, where $\delta_{ij}$ is the Kronecker delta function. This models the excitation of a single lattice site $(p,q)$ at frequency $\lambda$. In practice, we take a time-dependent function of the form $f_{\lambda}(t) = {\rm tanh} \left(  \frac{t}{10} \right) \exp(- i \lambda t)$ for $t \ge 0$ where a hyperbolic tangent function is used to gradually ramp up the field. 

Below we only consider  configurations corresponding to nontrivial topologies. In general, when  no gapless states exist (trivial local Chern number), we do not have localized conduction states; nontrivial edge  states are essential for this mechanism to work.

\subsection{Single Source}

\begin{figure}
\centering
   \includegraphics[scale = 0.4]{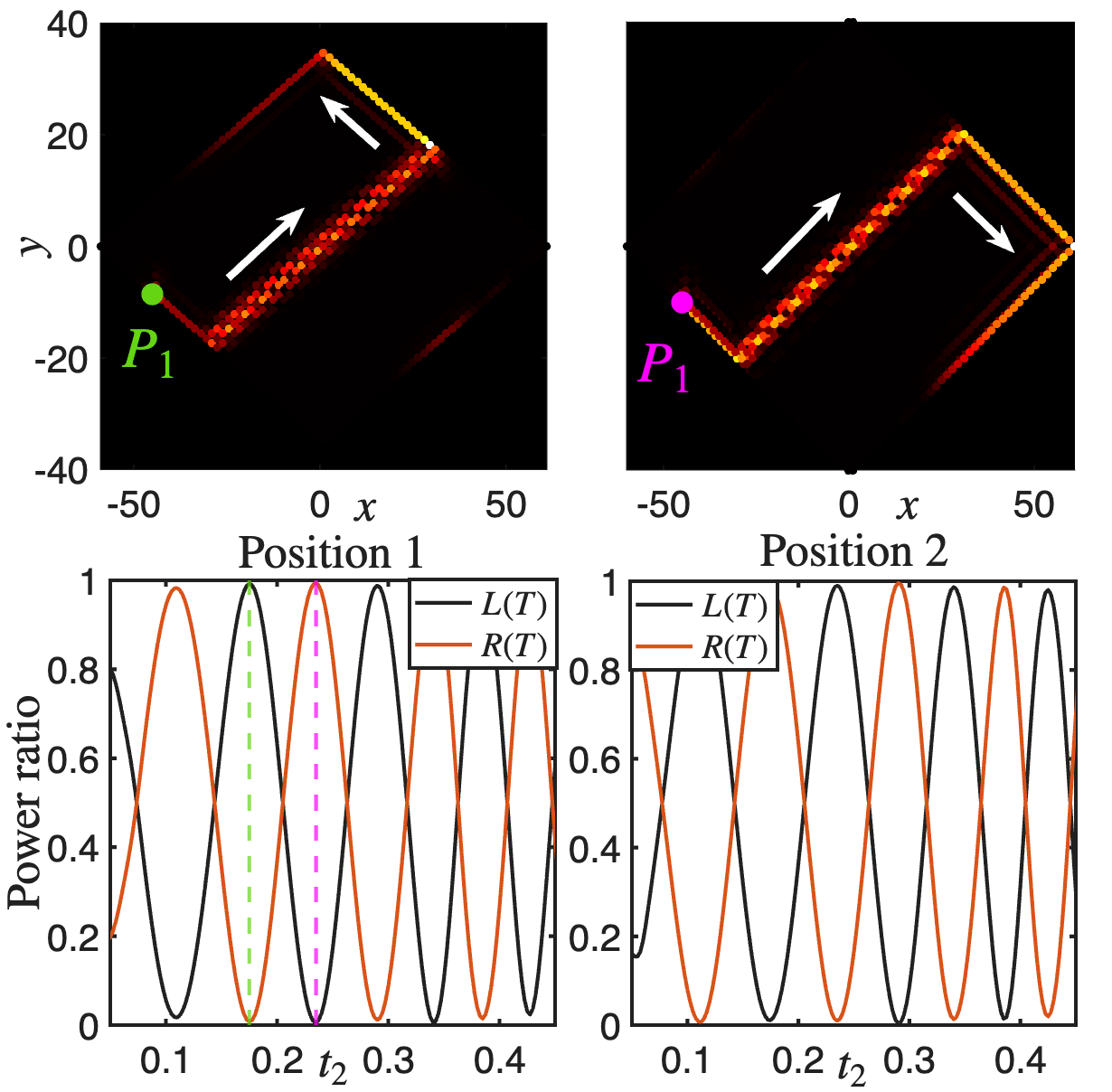}
    \caption{Single source antenna switching, as measured by power ratios (\ref{power_ratio}), is controlled by the $t_2$ coefficient in Eq.~(\ref{source_eqn}). Position 1  and Position 2  denote the location of the source antenna. The location of Position 1 is shown in the top row at time $t = T$; the location of Position 2 is shown in Fig.~\ref{2source_powerswitch}.  Two snapshots of $|{\bf c}(T)|$ are shown in the top row: switching to the left and right, respectively; snapshots of the Position 2 results are similar and not shown here.
     \label{singlesource_varyt2_summary}}
\end{figure}

A source is placed along the bottom-left edge $(m = -M)$ at Position 1 ($P_1$); see Figs.~\ref{singlesource_varyt2_summary} 
 and \ref{2source_powerswitch}. The energy emitted travels along the interface, in the ${\bf v}_1$ direction, until it reaches the top-right edge ($m = M$). At this junction, the field must turn left ($-{\bf v}_2$ direction) or right  (${\bf v}_2$ direction); there is no backscatter in the topological configuration. We wish to control the flow of energy out of this junction.

For a single source, we focus on adjusting the physical properties of the lattice and the source itself.
To track the flow of energy, we introduce the edge power ratios
\begin{align}
\label{power_ratio}
& {L}(t) = \frac{P_L}{P_L + P_R} , ~~~~ {R}(t) = \frac{P_R}{P_L + P_R} , \\ \nonumber
& P_L(t) = \sum_{n=-N}^{-1} | a_{M,n}|^2 + | b_{M,n}|^2 , \\ \nonumber
& P_R(t) = \sum_{n=1}^{N} | a_{M,n}|^2 + | b_{M,n}|^2 .
\end{align}
The quantities $L(t)$ and $R(t)$ measure
the relative amount of edge power  to the left ($n <0$) and right ($n > 0$) of the interface, respectively.  
We observe  that these quantities are more stable than tracking a single point, which can experience  temporary variations.

The first case  corresponds to adjusting the next-nearest neighbor modulus $t_2$, at fixed frequency $\lambda = 0$, and is shown in Fig.~\ref{singlesource_varyt2_summary}. Physically, this  corresponds to varying the strength (magnitude) of an applied magnetic field. 
The relationship between the parameters of the Haldane model and a magneto-optical lattice, for example, were discussed in an earlier work \cite{Ablowitz2024a}. We evolve Eq.~(\ref{source_eqn}) until a time $t = T$ when a steady-state  is observed and then measure the power ratio (\ref{power_ratio}).
As $t_2$ changes, we observe oscillation between light steered completely to the left or right. When the location of the source is switched to the other side of the interface (Position 2 ($P_2$) --see Fig. \ref{2source_powerswitch}), the direction of flow  changes. 
Snapshots in Fig.~\ref{singlesource_varyt2_summary}(top row) highlight the stark contrast between the different configurations.
Also notice that  it is possible  split the energy. For example, one can tune the parameter $t_2$ such that half the energy goes left and the other half goes right; see Fig.~\ref{singlesource_varyt2_summary}(bottom row) when $L,R \approx 0.5$.

\begin{figure}
\centering
   \includegraphics[scale = 0.425]{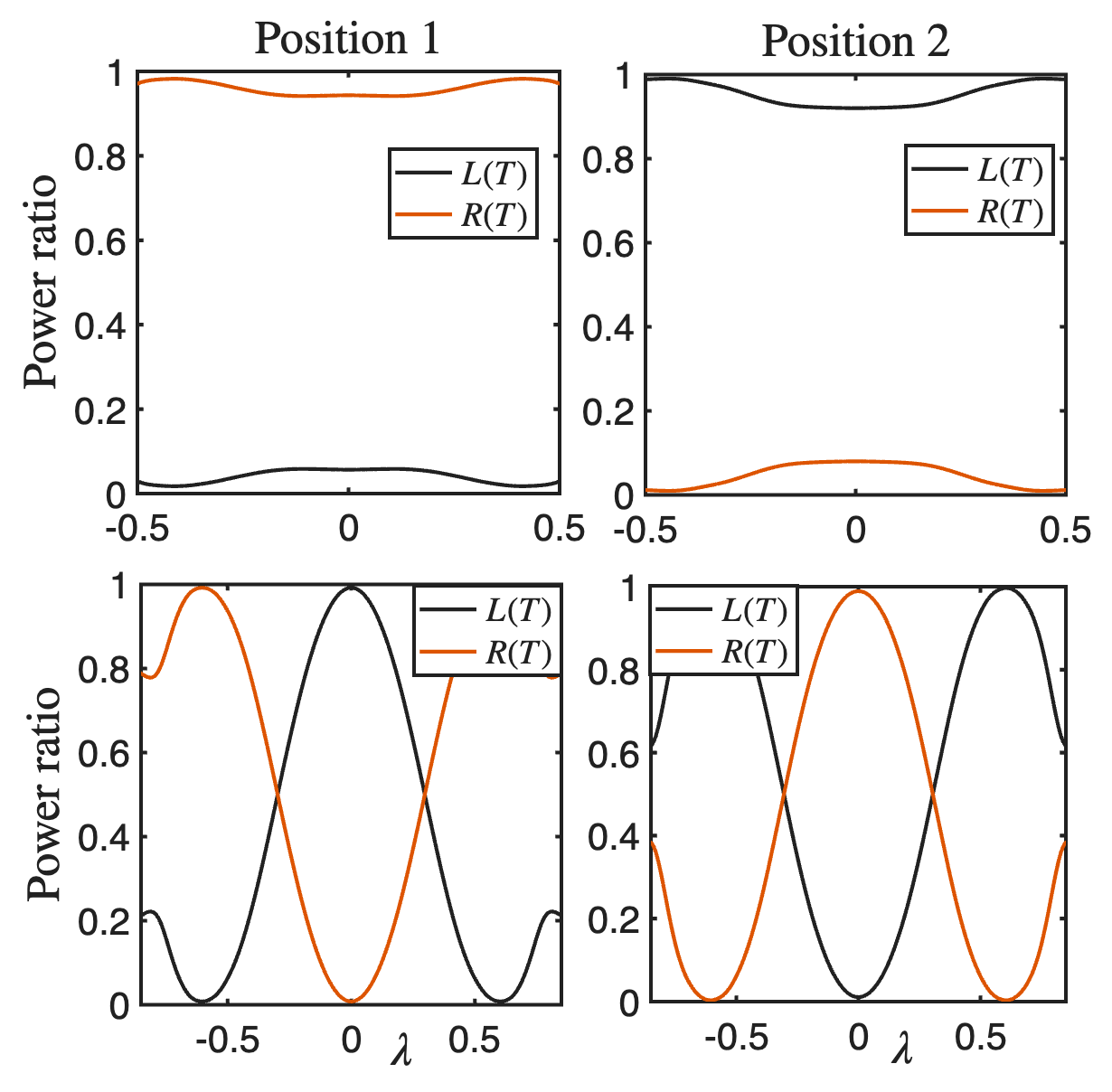}
    \caption{Single source antenna switching, as measured by power ratios (\ref{power_ratio}), is controlled by the frequency $\lambda$  in Eq.~(\ref{source_eqn}). Position 1 (left) and 2 (right) columns denote the location of the source antenna. Top row: $t_2 = 0.1$. Bottom row: $t_2 = 0.175$.
     \label{singlesource_varylambda_summary}}
\end{figure}

Next, the impact of the source frequency on routing is explored. Namely, we examine the switching dynamics as the frequency of the antenna source in Eq.~(\ref{source_eqn}) is adjusted for frequencies in the gap; the value of $t_2$ is fixed. The results highlighted in Fig.~\ref{singlesource_varylambda_summary} reveal an intimate dependency between flux strength $t_2$ and spectral value $\lambda$. In the first case, $t_2 = 0.1$, varying $\lambda$ has  little effect on the direction the energy flows; energy from Position 1 (2) tends to flow right (left).  On the other hand, at coupling strength $t_2 = 0.175$ the direction the energy
flows can effectively be controlled by adjusting $\lambda$. 
Physically, this indicates that, for a suitably configured  magnetic field, adjusting the frequency of the antenna source can be used to steer the flow of energy.

\begin{figure}
\centering
   \includegraphics[scale = 0.45]{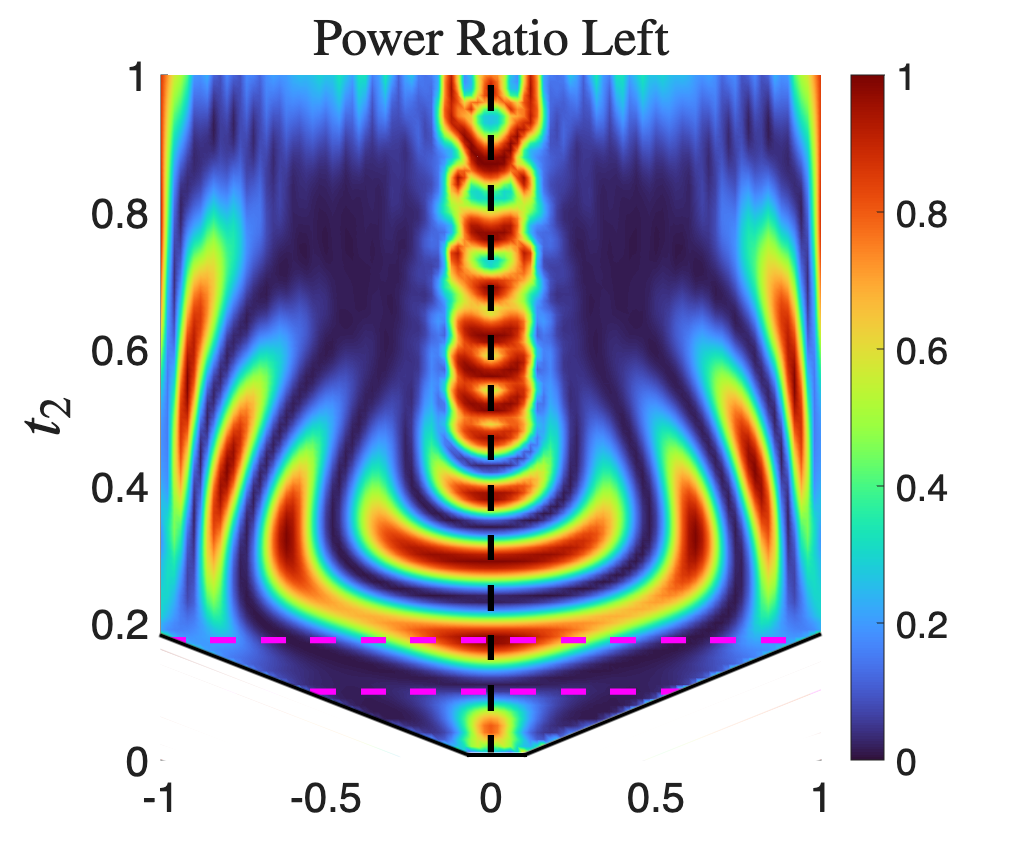}
      \includegraphics[scale = 0.45]{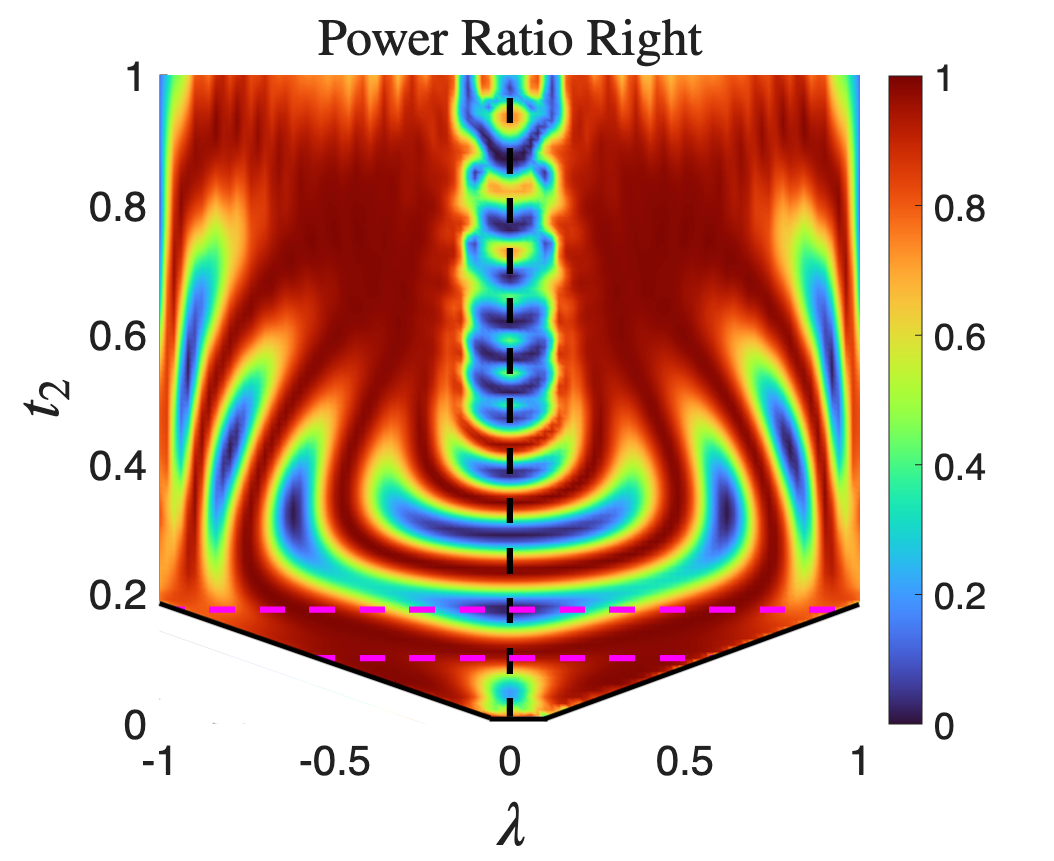}
    \caption{Single source antenna switching, as measured by left and right power ratios (\ref{power_ratio}) at time $T = 150$ for source antenna in Position 1. Other parameters are $t_1 = 1, \phi  = \pi/2$. The direction of switching depends, in a nontrivial manner, on both strength of magnetic flux $t_2$ and  source   frequency $\lambda$.  The black (magenta) dashed line corresponds to Fig.~\ref{singlesource_varyt2_summary} (Fig.~\ref{singlesource_varylambda_summary}).
     \label{2d_parmeter_source}}
\end{figure}

The nontrivial dependence on the model parameters is highlighted in Fig.~\ref{2d_parmeter_source}.
These results were obtained by placing a source antenna at Position 1 and evolving  Eq.~(\ref{source_eqn}) until a steady-state is reached. The results in Fig.~\ref{singlesource_varyt2_summary} and \ref{singlesource_varylambda_summary} correspond to vertical  (fixed $\lambda$, dashed black line) and horizontal  (fixed $t_2$, dashed magenta lines) cuts, respectively.
A  nontrivial dependence on the magnetic flux, $t_2$, and source frequency, $\lambda$ is revealed. 
 These figures indicate that  obtaining the results in Fig.~\ref{singlesource_varylambda_summary}  require fine tuning; i.e. there is sensitive dependence on the physical parameters. 
 The results for an antenna source in Position 2 effectively switch the left and right cases in Fig.~\ref{2d_parmeter_source} and are not shown here. We note that the bottom left and right corners of Fig.~\ref{2d_parmeter_source} are not included: they correspond to  spectral band  frequencies and do not support interface transmission; we discard them.

These results  suggest a nonlocal mechanism for the routing of  electromagnetic radiation. Rather than control the routing at the T-junction (where the light switches), an antenna located well away, is used to dictate the flow of energy. This formulation also has the benefit of being modular: an antenna source could be added to an existing multi-domain lattice and a router created. The sensitivity to physical parameters permits different splitting modalities: dependence on frequency (like Fig.~\ref{singlesource_varylambda_summary}(bottom row)) or lack thereof  (like Fig.~\ref{singlesource_varylambda_summary}(top row)).

Lastly, our results show good agreement with those measured in \cite{Tang2024}. In that work a magneto-optical lattice configuration, similar to the one considered here, was experimentally realized and numerically simulated. In particular, Figs.~\ref{singlesource_varyt2_summary} and \ref{singlesource_varylambda_summary} above correspond to Figs.~4 and 5 of \cite{Tang2024}, respectively, and capture the sinusoidal dependence on the magnetic field strength and source frequency.

To test the robustness of this system, we added on-site disorder to each lattice site. A normal distribution with mean zero and standard deviation of $\sigma = 0.1$  is taken. Relative to the topological transition (see Fig.~\ref{top_phasediagram}), one standard deviation corresponds to $\sigma / (3 \sqrt{3} t_2 \sin \phi) \approx 19 \%$. We view this as a substantial perturbation since two standard deviations is $38\%$ of the way to the point of topological transition.

\begin{figure}
\centering
   \includegraphics[scale = 0.29]{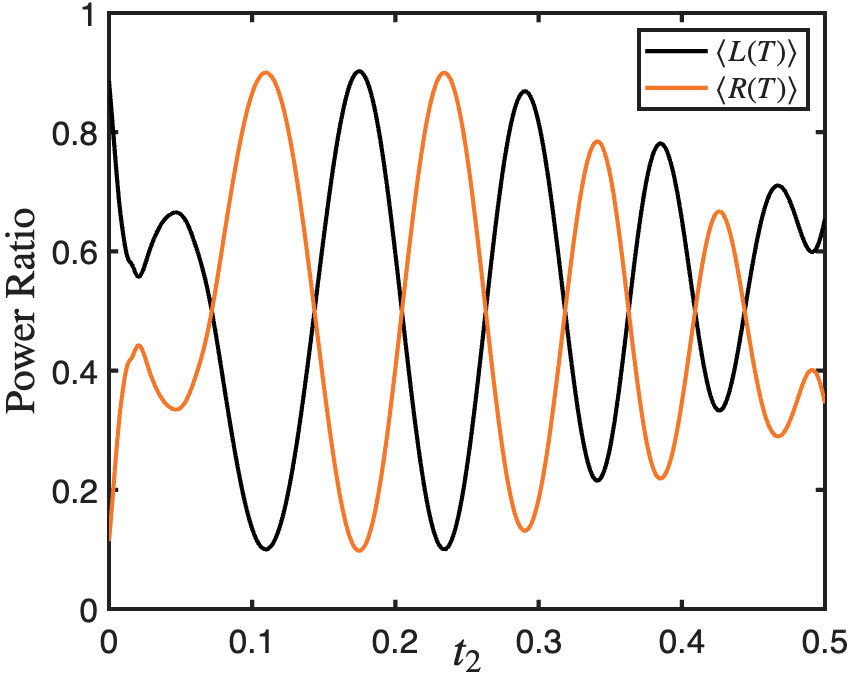}
      \includegraphics[scale = 0.29]{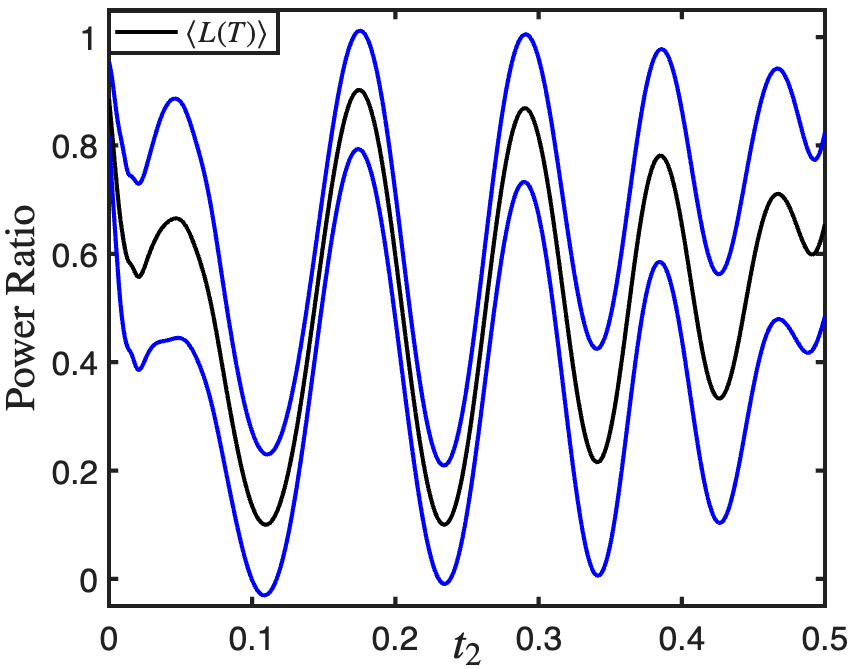}
    \caption{ Similar formulation and parameters to Position 1 result in Fig.~\ref{singlesource_varyt2_summary}(bottom-left). On-site normally distributed disorder has been added at each site with mean zero and standard deviation $0.1$. The average over 100 realizations is shown. (left) A left-right switching effect dependent on $t_2$ is still observed. (right) The average left ratio with its standard deviation bars included;  behavior to the right is reflected about the 50\% power ratio line.
     \label{1d_disorder_test}}
\end{figure}

Average results for the one-antenna Position 1 routing setup (see Fig.~\ref{singlesource_varyt2_summary}) for 100 realizations are shown in Fig.~\ref{1d_disorder_test}. Overall, we still observe a $t_2$-dependent switching effect despite significant disorder.
What is clear is the asymmetry is not a stark as the clean case. For the most prominent peaks, we observe a 90/10-splitting ratio whereas in the unperturbed cases shown in Fig.~\ref{singlesource_varyt2_summary} splitting ratios approaching 99/1 are observed.

Standard deviation bars are shown in Fig.~\ref{1d_disorder_test}(right) to show  variation from the mean. We note that while these suggest ratio values of greater than 1 or less than 0, that cannot actually occur (by definition). In these cases, the deviation is asymmetry about the average. 
These stability results indicate that the routing of light in topological insulators is robust even in the presence of significant uncorrelated disorder.

\subsection{Two Sources}

Next, we consider an arrangement with two different antenna sources along the bottom edge ($m = - M$). The benefit of using two sources, rather than just one, is that for a general set of parameters $t_2,\lambda$, we can tune the switching effect. This technique is an extension of the approach developed in \cite{Ablowitz2024}. Namely, by tuning the amplitude and phase  of the sources, we can control the direction light flows. Hence, this represents an on demand,  nonlocal (control at a distance) approach to controlling the routing. 

\begin{figure}
\centering
   \includegraphics[scale = 0.3]{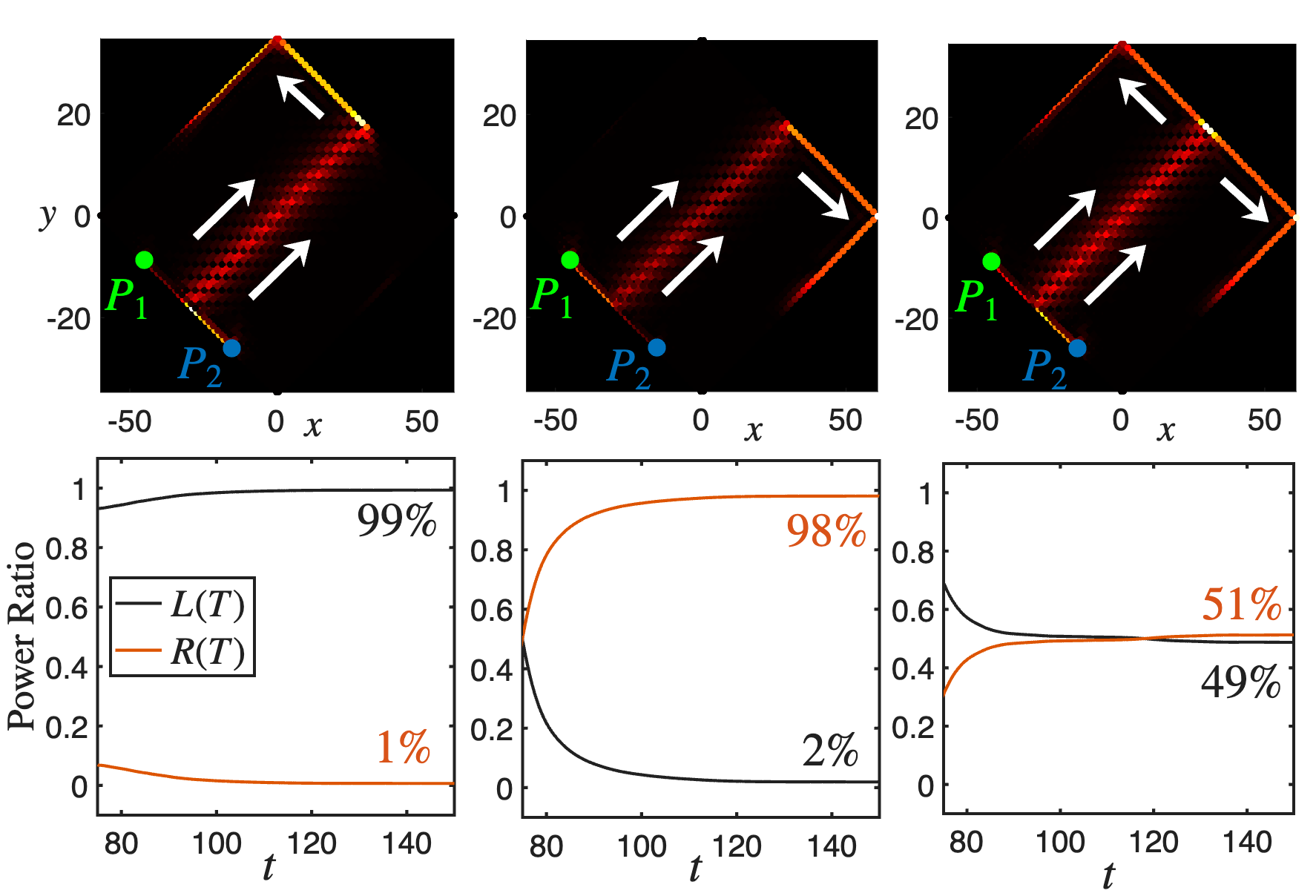}
    \caption{Summary of the two-source topological switching effect. Two antenna sources (green and blue dots) inject light at frequency $\lambda = 0$; other parameters are the same as Fig.~\ref{eigenenergy_2d_compare}(a). Top row, left-to-right: snapshot of a left, right, and even switch at time $T  = 150$. Bottom row: Corresponding power ratios (\ref{power_ratio}) with  power ratio  value at final time $t = T$ indicated.
 \label{2source_powerswitch}}
\end{figure}

The setup consists of modifying the source function in Eq.~(\ref{source_eqn}) to excite two lattice sites: Position 1 {\it and} Position 2. The location of the sources is indicated in Fig.~\ref{2source_powerswitch}(top row). Both sites are excited at the same frequency of the form
\begin{equation}
\label{two_source_force}
{\bf f}(t) \rightarrow  \left( c_1 \delta_{1i} + c_2 \delta_{2i} \right)f_{\lambda}(t) .
\end{equation}
where $f_{\lambda}(t) $ is defined below Eq.~(\ref{source_eqn}) and $\delta_{1i},\delta_{2i}$  denote the $P_1, P_2$ source locations, respectively.

A  transfer matrix approach that was developed in \cite{Ablowitz2024} can be used to tune the switch. Namely, for a fixed set of parameters ($t_2$ and $\lambda$), one can determine the coefficients $c_1,c_2$ in (\ref{two_source_force}) for a desired output.
Consider two output sites $b_{L} $ and $b_{R}$, which are located to the left and right, respectively, of the switching junction. For the results below, we chose $b_L = b_{M,-15}$ and $b_R = b_{M,15}$.
This is incorporated into the linear system
\begin{equation}
\label{transfer_matrx}
\mathcal{T} {\bf c} = {\bf b} ,
\end{equation}
where ${\bf c} = (c_1, c_2)^T$ and ${\bf b}  = (b_L , b_R)^T$. What this equation says is the transfer matrix $\mathcal{T}$ maps a set of weights ${\bf c}$ to some output ${\bf b}$. In practice, we actually want to invert this. That is, given a desired output ${\bf b}$, we want to determine the appropriate coefficients ${\bf c} = \mathcal{T}^{-1} {\bf b}$.

To determine the transfer matrix $\mathcal{T}$, we make two runs. First, we excite only the source in Position 1 ($c_1 = 1, c_2 = 0$), allow the state to reach a steady-state, and measure the values at $b_L$ and $b_R$; we denote these by $b_{L}^{(1)}$ and $b_{R}^{(1)}$. We repeat this process, exciting from the second source position ($c_1 = 0, c_2 = 1$) and measure the outputs, given by $b_L^{(2)}$ and $b_R^{(2)}$. The elements of the transfer matrix are
\begin{equation*}
\mathcal{T} = \begin{pmatrix} b_L^{(1)} & b_L^{(2)} \\ b_R^{(1)} & b_R^{(2)}  \end{pmatrix} .
\end{equation*}
Observe that $\mathcal{T} \begin{pmatrix} 1 \\ 0 \end{pmatrix}  = \begin{pmatrix} b_L^{(1)} \\ b_R^{(1)} \end{pmatrix}$ and  $\mathcal{T} \begin{pmatrix} 0 \\ 1 \end{pmatrix}  = \begin{pmatrix} b_L^{(2)} \\ b_R^{(2)} \end{pmatrix}$. 

Typical results are shown in Fig.~\ref{2source_powerswitch}. From left-to-right, the results show left, right, and even split switching. In the case of left or right switching, 98-99\% of power moves in the preferred direction. To find the appropriate coefficients, we consider outputs ${\bf b} = (1,0)^T$, $(0,1)^T$, and $ (0.5,0.5)^T$, respectively. For these parameters, the corresponding coefficients obtained by solving (\ref{transfer_matrx}) are: ${\bf c} \approx (0.1429,0.5612)^T, (-0.5593,0.1435)^T$, and $(-0.2082,0.3523)^T$.

Similar to the one source case considered in Fig.~\ref{1d_disorder_test}, we added on-site disorder and examined the effect on the two source switching protocol. To maintain consistency, the same realization of disorder was used in each case. Despite significant uncorrelated disorder, we reproduce routing ratios comparable to those shown in Fig.~\ref{2source_powerswitch}. The transfer matrix approach in Eq.~(\ref{transfer_matrx}) is adaptive and the coefficients $c_1,c_2$ change with the introduction of disorder. This represents a more systematic approach to tailoring the routing of the energy.

A comparison between the one and two source approaches is appropriate. The one source construction is simpler  in the sense that it only requires one antenna input. On the other hand, it requires tuning physical properties such as the magnetic flux strength or the frequency, whose dependencies appears  to be highly nontrivial (see Fig.~\ref{2d_parmeter_source}). The primary benefit of two source configuration is that switching can be achieved for arbitrary parameters, on demand. However, in general  it requires adjusting the amplitude and phase delay at both source locations. Both approaches are capable of directing the flow of energy: completely left, completely right, or some splitting, in a nonlocal fashion.

\section{Conclusion}

In this work we have developed a two-region tight-binding Haldane-type model that supports topologically protected interface states. By adjusting the physical parameters in the model; e.g. corresponding to magnetic field strength or frequency, it is possible to control the flow of energy at a junction; effectively creating a switch. It is found that it is possible to interfere two antennas sources and direct the flow of light. Due to the topological nature of this system, this construction is robust to perturbations. All of these approaches represent a {\it nonlocal} switching mechanism: namely, the flow of energy is not controlled at the switching junction, but rather by source antenna(s) located well away.

Indeed the Haldane model is a fundamental model and has been realized in a wide variety of physical systems; e.g. \cite{Goldman2016,Jotzu2014,Lannebere2018}. As a result the results in this paper can be expected to apply  broadly.

This mechanism represents an intriguing
 feature of Chern 
 topological insulators
 that has the potential to be utilized in future photonic devices. An interesting future direction is to stack several of these regions together. That is, given a single input can energy be steered on demand to any number of output ports? The robust nature of topological insulators indicates that such a device would be quite efficient due to minimal scattering loss.
 Furthermore, since  this effect has been observed in both magneto-optical \cite{Tang2024} and Floquet photonic lattices \cite{Ablowitz2024}, it can be expected that this phenomena will  be found in other
 Chern insulator systems too.

\section{Acknowledgements}
This work was supported by the Air Force Office of Scientific Research (AFOSR) under  Grant No. FA9550-23-1-0105 (J.T.C.) and NSF under Grant No. 2306290 (M.J.A.).

\appendix

\section{Hamiltonian Representation}
\label{Hamil_define_sec}

The governing equations in Eqs.~(\ref{a_eqn})-(\ref{b_int_eqn}) are encoded in the Hamiltonian matrix $H$, given in Eqs.~(\ref{compact_time_ind}), (\ref{spec_localizer}), and (\ref{source_eqn}). Due to its importance throughout this paper, here we write down the matrix $H$ explicitly. For simplicity, we set the mass term $\mathcal{M} $ to zero; it is  simple to re-incorporate it from the formulation below when it is nonzero.

Consider a  honeycomb lattice system of lattice sites $a_{mn} ,b_{mn}$ for $m = - (M+1) , \dots, M+1$ and $n = - (N+1) , \dots, N+1$, with an interface  located at $n = 0$. This implies a system of  $2 \times (2M+1) \times (2N+1)$ unknowns. Zero boundary conditions (\ref{zero_BCs}) are applied along the perimeter.
We choose a bipartite storage, i.e. ${\bf v} = [{\bf a} | {\bf b}]^T$, where
\begin{align*}
{\bf a} & = [{\bf a}_{-N-1} | {\bf a}_{-N} |  \cdots |  {\bf a}_{N+1}]^T \\
{\bf b} & = [{\bf b}_{-N-1} | {\bf b}_{-N} |  \cdots |  {\bf b}_{N+1}]^T ,
\end{align*}
and
\begin{align*}
{\bf a}_{n} & = [a_{-M-1,n} | a_{-M,n} |  \cdots |  a_{M+1,n}]^T \\
{\bf b}_{n} & = [b_{-M-1,n} | b_{-M,n} |  \cdots |   b_{M+1,n}]^T .
\end{align*}

The Hamiltonian matrix is structured as
\begin{equation}
\label{Hamiltonian_form}
H = \begin{bmatrix} H_{aa} & H_{ab} \\ H_{ba} & H_{bb} \end{bmatrix} ,
\end{equation}
where  $H_{ij}$ indicates the coupling of the $j$ sites with the $i$ sites. Each component of  Eq.~(\ref{Hamiltonian_form}) is a block matrix. For interactions of the same species (e.g. $a$-site interactions with other $a$-sites),  the interface induces the structure
\begin{equation}
H_{aa}  = 
  \left[ \begin{array}{ *{5}{c} }
    \multicolumn{2}{c}
      {\raisebox{\dimexpr\normalbaselineskip+.4\ht\strutbox-2.25\height}[0pt][0pt]
        {\scalebox{1.5}{$\mathcal{H}_N$}}} & \mathcal{O}& \mathcal{O} & \mathcal{O} \\
          & & \mathbb{C}& \mathcal{O} & \mathcal{O}  \\
 \mathcal{O}  &\mathbb{C}^\dag & \mathbb{D }  & \mathbb{C}^* & \mathcal{O}  \\
  \mathcal{O}  & \mathcal{O}  & \mathbb{C}^T  &    \multicolumn{2}{c}
      {\raisebox{\dimexpr\normalbaselineskip+.4\ht\strutbox-2\height}[0pt][0pt]
       {\scalebox{1.5}{$\mathcal{H}_N^*$}}} \\
        \mathcal{O} &  \mathcal{O} &  \mathcal{O}  \\
  \end{array} \right] , 
\end{equation}
\begin{equation}
H_{bb}  = 
  \left[ \begin{array}{ *{5}{c} }
    \multicolumn{2}{c}
      {\raisebox{\dimexpr\normalbaselineskip+.4\ht\strutbox-2.25\height}[0pt][0pt]
        {\scalebox{1.5}{$\mathcal{H}_N^*$}}} & \mathcal{O}& \mathcal{O} & \mathcal{O} \\
          & & \mathbb{C}^*& \mathcal{O} & \mathcal{O}  \\
 \mathcal{O}  &\mathbb{C}^T & \mathbb{D }  & \mathbb{C} & \mathcal{O}  \\
  \mathcal{O}  & \mathcal{O}  & \mathbb{C}^\dag  &    \multicolumn{2}{c}
      {\raisebox{\dimexpr\normalbaselineskip+.4\ht\strutbox-2\height}[0pt][0pt]
       {\scalebox{1.5}{$\mathcal{H}_N$}}} \\
        \mathcal{O} &  \mathcal{O} &  \mathcal{O}  \\
  \end{array} \right] , 
\end{equation}
where $^*$ denotes complex conjugation, $^T$ is the transpose, and $^\dag$ is complex conjugation transpose. 
Define the  matrix  $\mathcal{H}_{N}$ in terms of Kronecker products
\begin{equation}
 \mathcal{H}_{N} = \mathbb{I}_N \otimes \mathbb{D} + \mathbb{U}_{N} \otimes \mathbb{C} + \mathbb{L}_N \otimes \mathbb{C}^\dag ,
\end{equation}
where $\mathbb{I}_N$ is the $N \times N$ identity matrix, and  the  $N \times N$ upper diagonal matrix 
\begin{equation}
\mathbb{U}_N = \begin{bmatrix}
0 & 1 \\
& 0 & 1 \\
 & & \ddots & \ddots  \\  
 & & &&  & 1 \\
  & & &&  & 0
\end{bmatrix}
\end{equation}
 and the lower triangular $\mathbb{L}_N = \mathbb{U}_N^T$, which mirrors $\mathbb{U}_N$ about the main diagonal. 
Define the $(2M+1) \times (2M+1)$ matices
\begin{equation}
\mathbb{D} = \begin{bmatrix} 
0 & t_2 e^{i \phi} & \\
t_2 e^{-i \phi} & 0 & t_2 e^{i \phi} \\
 & t_2 e^{-i \phi} & \ddots & \ddots \\
& & \ddots \\ \\
 & & &  & & t_2 e^{i \phi} \\
  & & & & t_2 e^{-i \phi}  & 0
\end{bmatrix} ,
\end{equation}
and
\begin{equation}
\mathbb{C} = \begin{bmatrix} 
t_2 e^{- i \phi} &  & \\
t_2 e^{i \phi} & t_2 e^{-i \phi} &  \\
 & t_2 e^{i \phi} & \ddots & \\
& & \ddots \\ 
 & & &  & & & t_2 e^{-i \phi} \\
  & & & &  & & t_2 e^{i \phi} 
\end{bmatrix} .
\end{equation}
Note that both $H_{aa}$ and $H_{bb}$ are Hermitian matrices, as they must be for $H$ to be Hermitian.

Next, define the inter-species coupling matrices $H_{ab}$ and $H_{ba}$. To begin, define the nearest neighbor coupling matrix
\begin{equation}
\mathbb{N} = \begin{bmatrix} 
t_1 &  & \\
t_1 & t_1&  \\
 & t_1 & \ddots & \\
& & \ddots \\ 
 & & &  & & & t_1 \\
  & & & &  & & t_1
\end{bmatrix} .
\end{equation}
Since the nearest neighbor coupling is assumed to be independent of magnetic field, this takes the relatively simple form
\begin{equation}
H_{ab} = \mathbb{I}_{2N+1} \otimes \mathbb{N}  +  \mathbb{L}_{2N+1} \otimes (t_1 \mathbb{I}_N)  .
\end{equation}
Lastly, $H_{ba} = H_{ab}^\dag$; this is necessary for $H$ to be Hermitian.

\end{document}